\documentclass[aps,pra,twocolumn]{revtex4}
\usepackage{mathtools,amssymb,lipsum}
\usepackage{graphicx}

\usepackage{xcolor,ulem}

\begin{document}

\title{Enantioselective manipulation of single chiral nanoparticles using optical tweezers } 
\author{R. Ali}
\email[]{r.ali@if.ufrj.br}
\affiliation{%
Instituto de F\'isica, Universidade Federal do Rio de Janeiro, Caixa Postal 68528, Rio de Janeiro, RJ, 21941-972, Brasil
}
\author{F. A. Pinheiro}
\affiliation{%
Instituto de F\'isica, Universidade Federal do Rio de Janeiro, Caixa Postal 68528, Rio de Janeiro, RJ, 21941-972, Brasil
}
\author{ F. S. S. Rosa}
\affiliation{%
Instituto de F\'isica, Universidade Federal do Rio de Janeiro, Caixa Postal 68528, Rio de Janeiro, RJ, 21941-972, Brasil
}
\author{ R. S. Dutra}
\affiliation{LISComp-IFRJ, Instituto Federal de Educa\c{c}\~ao, Ci\^encia e Tecnologia, Rua Sebasti\~ao de Lacerda, Paracambi, RJ, 26600-000, Brasil}
\author{P. A. Maia Neto}
\affiliation{%
Instituto de F\'isica, Universidade Federal do Rio de Janeiro, Caixa Postal 68528, Rio de Janeiro, RJ, 21941-972, Brasil
}

\date{\today}

\begin{abstract}

We put forward an enantioselective method for  chiral nanoparticles using optical tweezers.  We demonstrate that the optical trapping force in a typical, realistic optical tweezing setup with circularly-polarized trapping beams is sensitive to the chirality of core-shell nanoparticles, allowing for efficient enantioselection. It turns out that  the handedness of the trapped particles can be selected by choosing the appropriate circular polarization of the trapping beam.
The chirality of each individual trapped nanoparticle can be characterized by measuring the rotation of the equilibrium position 
under
 the effect of a transverse Stokes drag force. 
 We  show that
 the chirality of the shell gives rise to an additional twist,  
leading to a strong enhancement of the optical torque driving the rotation. 
Both  methods are shown to be robust against variations of  size and material parameters, demonstrating that 
they are particularly useful in (but not restricted to)  several situations of practical interest in chiral plasmonics, where enantioselection and characterization of single chiral nanoparticles, each and every with its unique handedness and optical properties, are in order.
In particular, our method could be employed to unveil the chiral response arising from disorder 
in individual  plasmonic raspberries, 
synthesized by close-packing a large number of metallic nanospheres around a dielectric core. 
\end{abstract}

\maketitle

The concept of chirality, introduced by Lord Kelvin to designate any geometrical object or ensemble of points that
lack mirror symmetry, is ubiquitous in Nature~\cite{wagniere}. 
Chirality is  essential in several nanotechnological applications~\cite{zhang2005}. 
Plasmonic structures provide a promising material platform for artificially designed chiral systems and applications ~\cite{cpreview1,cpreview2,cpreview3}. 
One of the major goals of the emerging field of chiral plasmonics is to  tailor optical chiral responses 
orders of magnitude larger than those of natural chiral materials. 

Indeed, new classes of chiral systems composed of plasmonic nanoparticles have been recently developed. Examples include achiral metallic particles arranged in chiral geometries, such as helices~\cite{fan2010} or random configurations~\cite{pinheiro2017}, chiral metallic nanocrystals~\cite{fan2012}, DNA-assembled plasmonic nanostructures~\cite{kuzyk2012,yan2012}, chiral plasmonic particles assembled on scaffolds~\cite{merg2016}, and core-shell plasmonic spheres~\cite{wu2015,liu2013}. 
However, the handedness and geometrical chirality of many of these artificial plasmonic structures, which ultimately govern the optical chiral response, are not {\it a priori} known after the nanofabrication process. Besides, many applications involve single, isolated chiral nanoparticles, each one with unique, unkwown handedness, geometry and optical properties. In these cases, traditional probes of chirality, such as rotatory power and circular dichroism \cite{dich}, are expected to fail in probing single-particle properties as they typically provide an average chiral response of these structures in solution.

\begin{figure}
\includegraphics[width = 3.4 in]{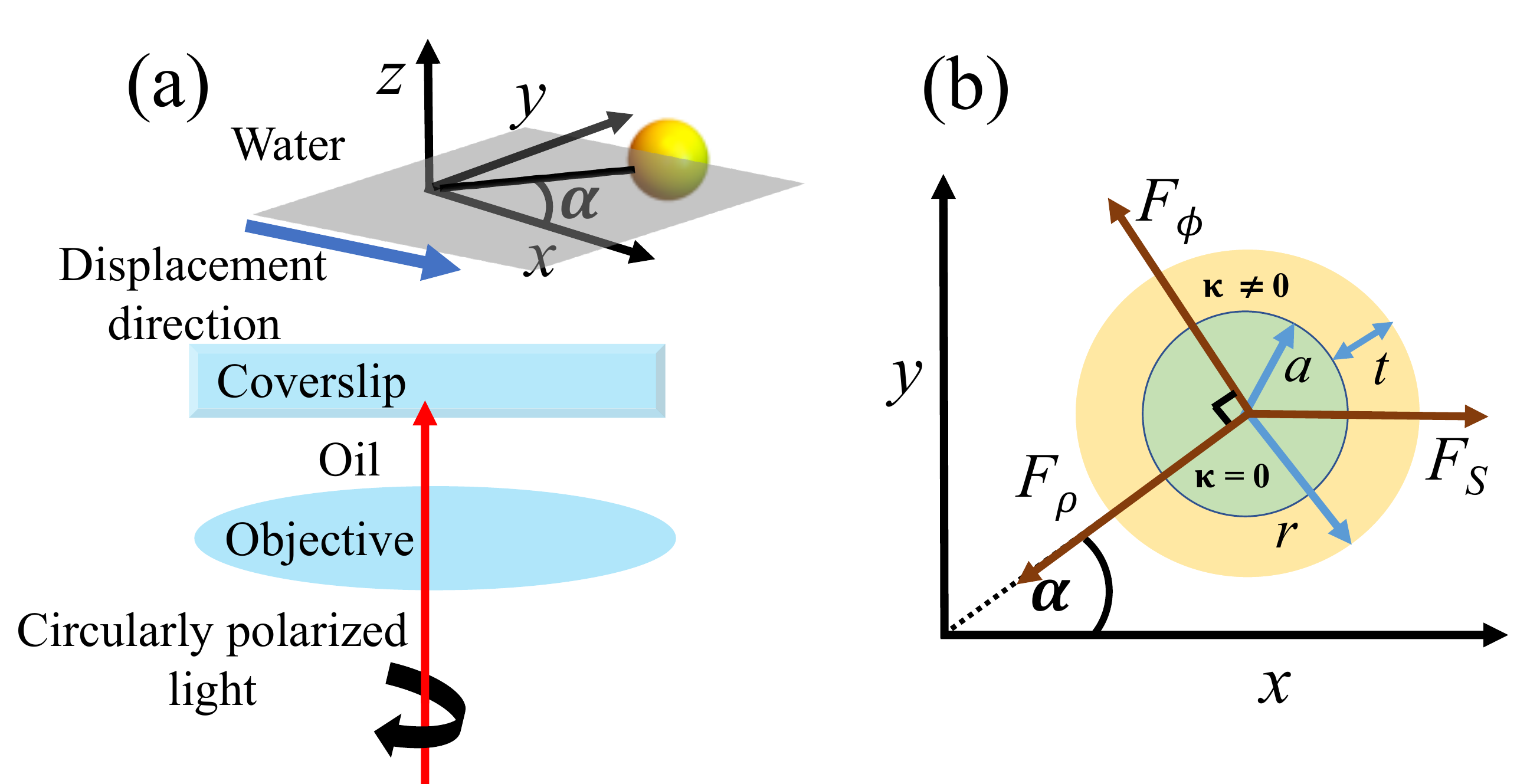}
\caption{ (a) Schematic representation 
of the proposed enantioselection with optical tweezers. A circularly-polarized Gaussian laser beam is focused into a spot at the focal plane of a high-NA objective, trapping
core-shell nanospheres in the aqueous solution. The handedness of the chiral shell allowing for trapping is selected by choosing the appropriate circular polarization. In order to provide for further 
enantioselectivity and characterization of the chirality parameter, the trapped nanosphere is displaced off-axis by driving the sample along the $x$ direction with a constant speed. (b) 
The resulting Stokes drag force $F_S$ is represented alongside the cylindrical optical force components $F_{\rho}$ and  $F_{\phi}.$
The latter accounts for the OAM transferred to the nanosphere. The equilibrium position is rotated around the $z$-axis by the angle $\alpha.$} 
 \label{setup}
\end{figure}

Here we propose using optical tweezers (OT)~\cite{Ashkin1986} with a single circularly-polarized trapping beam to select and 
probe the handedness of single, isolated chiral nanoparticles.
By applying for the first time optical tweezers to plasmonic particles, we demonstrate all-optical enantioselection and manipulation of an important class of materials in chiral plasmonics, namely core-shell metallic chiral nanoparticles, which are an excellent model for metal-organic nanoparticles with strong optical activity~\cite{wu2015}.
In our proposal,  sketched in Fig. \ref{setup}(a), 
a circularly-polarized Gaussian laser beam propagating along the $z$-direction reaches an oil immersion high-NA objective that focuses the beam into a difraction-limited spot in the sample beyond the coverslip.
The sample contains 
chiral nanoparticles dispersed in water. 
We show that optical trapping can only occur for particles  with the handedness 
selected by the choice of the trapping beam circular polarization,
thus allowing for enantioselectivity. 

Our proposal is  simpler than previous
 optical chiral resolution methods~\cite{li2007,spivak2009,li2010,cameron2014,durand2013,bradshaw2014,wang2014,hayat2015,alizadeh2015,durand2016,chen2016,dionne2016,zhang2017,acebal2017,ho2017,cao2018,dionne2018}, including those which have been  experimentally implemented~\cite{hernandez2013,tkachenko2014,donato2015,tkachenko2014b,dionne2017,Schnoering2018,krevets2019,Nker2019}. 
 More importantly, it allows for a quantitative measurement of the chirality parameter $\kappa$ of single, individual nanoparticles, an important task that to the best of our knowledge has never been achieved so far~\cite{dionnereply}. For that purpose,  
 we drive the sample laterally and measure the resulting rotation angle $\alpha$ of the equilibrium position shown in Fig.~1.
As illustrated by Fig.~1(b), the rotation around the laser beam axis $z$ is a consequence of the optical torque produced by the azimuthal optical force component 
$F_{\phi} $
exerted on the trapped microsphere when the latter is displaced off-axis 
 by the Stokes drag force $F_S.$ In other words, 
 $F_{\phi} $ accounts for the partial transfer of optical angular momentum to the trapped particle and its 
 sign is controlled by the optical helicity.
 
More specifically,  the rotation angle at equilibrium $\alpha$ is found from $\tan\alpha=F_{\phi}/|F_{\rho}|,$
 where $F_{\rho}<0$ is the restoring radial optical force also  depicted in Fig.~\ref{setup}(b).
 Due to the symmetry of the beam profile, both $F_{\phi}$ and $F_{\rho}$ are independent of the angular position of the sphere.

 For achiral homogenous microspheres, the angle of rotation $\alpha$ was measured and compared with theory~\cite{Diniz2019}. 
It turns out, as we show below, that  for chiral shells $\alpha$ is considerably larger than the typical values measured in~\cite{Diniz2019}, indicating that our proposal can indeed be implemented experimentally. In addition, we show that $\alpha$ is strongly dependent on the shell chirality parameter $\kappa$, allowing for a determination of $\kappa$ from the measurement of $\alpha.$

  A general formalism of optical torques on chiral particles was developed in Ref.~\cite{genet2015}.
 Here we compute the optical torque and the resulting rotation angle by extending the Mie-Debye with Spherical Aberration (MDSA) theory of optical tweezers \cite{MaiaNeto2000,Mazolli2003,Viana2007,Dutra2007} to encompass the case of nanospheres with a chiral coating.
 Our theoretical construction consists of three steps: 
  (i) start from a realistic non-paraxial description of the highly-focused trapping beam, taking into account the spherical aberration introduced by refraction at the interface between the coverslip and the sample;
  (ii) derive the scattered field by taking the representation for the focused beam 
 obtained in (i)  as the incident field on a sphere coated with a chiral layer; and (iii) integrate the resultant Maxwell stress tensor over an apropriate Gaussian surface  to derive the optical force. An alternative theory of optical tweezers for  homogeneous chiral particles and clusters based on the T-matrix formalism has recently appeared \cite{Patti2019}.
 
 Step (i) is based on the non-paraxial Debye model of a focused beam, later extended in Ref.~\cite{RichardsWolf} to account for polarization. We also introduce the spherical aberration 
 effect by following Ref. \cite{Torok1995}. Step (ii) requires an extension of MDSA involving a description 
 of chiral media. 
 In their most simple and useful instance, the constitutive equations \cite{Post} mixing the electric field $\bf{E}$, the electric displacement $\bf{D}$, the magnetic field  $\bf{B}$ and the auxiliary field ${\bf H}$ for a chiral medium are given by \cite{lindell, lakhtakia}
\begin{equation}
\begin{aligned}
{\bf D}({\bf r},\omega)=\epsilon_0\epsilon(\omega) {\bf E}({\bf r},\omega) +i \sqrt{\epsilon_0\mu_0} \,\kappa(\omega) {\bf H}({\bf r},\omega) ,\\ 
{\bf B}({\bf r},\omega)= -i \sqrt{\epsilon_0\mu_0}\,\kappa(\omega){\bf E}({\bf r},\omega) +\mu_0 \mu(\omega) {\bf H}({\bf r},\omega) ,
\label{ChiralFields}
\end{aligned}
\end{equation} 
where $\epsilon$ and $\mu$ are the relative permittivity and permeability of the medium  ($\epsilon_0$ and $\mu_0$ are the absolute vacuum permittivity and permeability), and $\kappa$ is the (relative) chirality parameter. 

We first write the incident focused beam as a superposition of plane waves, and then 
solve for the scattering of each plane wave component
by following the method of Ref.~\cite{Bohren1974}
and imposing the appropriate boundary conditions at each interface of the coated nanosphere. The total scattered field is finally obtained by superposing all scattered components. Finally, the optical force ${\bf F}$ is derived by integration of the stress tensor [step (iii)]. 
Momentum conservation allows one to take a spherical Gaussian surface $S(R)$ at infinity:
 \begin{eqnarray}
 {\bf F} = \lim_{R\rightarrow\infty}\left[ -\frac{ R}{2} \int_{S(R)} \! d\Omega \, {\bf r} \left( \epsilon_0\epsilon_w E^2 + \mu_0 H^2 \right)\right],
 \end{eqnarray}
 where  $\epsilon_w =n_w^2$ and $n_w$ is the water refraction index.
 
 The resulting optical force is written as a multipole series in terms of the
coefficients 
\begin{eqnarray}
 G^{(\sigma)}_{\ell m}(\rho,z) =  \int_0^{\theta_0} \hspace{-15pt }&& d\theta \sin\theta \sqrt{\cos\theta}\,e^{-\gamma^2\sin^2\theta}
  d_{m,\sigma}^{\ell}(\theta_w) T(\theta)   \nonumber \\
  \label{Gjm}
 &&\times \; e^{i\Phi_{w}(\theta)} J_{m-\sigma}(k\rho\sin\theta)\,e^{i k_{w}\cos\theta_{w} z },
 \end{eqnarray}
 which are functions of the sphere position (with respect to the paraxial focus) in cylindrical coordinates $(\rho,z),$
 with $\ell,m$ representing the multipole order. 
Explicit expressions for the  cylindrical components of ${\bf F}$ are given in the supplementary material \cite{supp_}.
 We take $\sigma=\pm 1$ for left-handed/right-handed circular polarizations. 
 The integration variable $\theta$ represents the angle between each incident propagation direction in the glass medium (index $n_g$) and the $z$-axis. 
 The integration limit is defined by the objective numerical aperture (NA): $\theta_0=\arcsin(\mbox{NA}/n_g).$
 The parameter $\gamma$ represents the  ratio of the objective focal length to the Gaussian beam waist at the objective entrance port 
 (but note that the focused beam is not described by a paraxial Gaussian model). 
   $T(\theta)$ is the Fresnel transmission amplitude for refraction at the interface between the glass coverslip and the aqueous solution. 
   Such refraction also introduces the spherical aberration phase correction $\Phi_{w}(\theta).$ 
   Explicit expressions are  given in \cite{supp_}.
   $d_{m,m'}^{\ell}(\theta_w)$ are the Wigner rotation matrix elements \cite{Edmonds}
   evaluated at the angle $\theta_w=\arcsin(n_g \sin\theta/n_w)$ in the aqueous solution and $J_{m}$ are the cylindrical Bessel functions of integer order $m$ \cite{DLMF25.12}.
   Finally,   $k=2\pi n_g/\lambda_0$
   and $k_w=n_wk/n_g$ are the wavenumbers in the glass and aqueous solution, respectively,
    where $\lambda_0$ is the vacuum wavelength.

\begin{figure}
\includegraphics[width = 3.4 in]{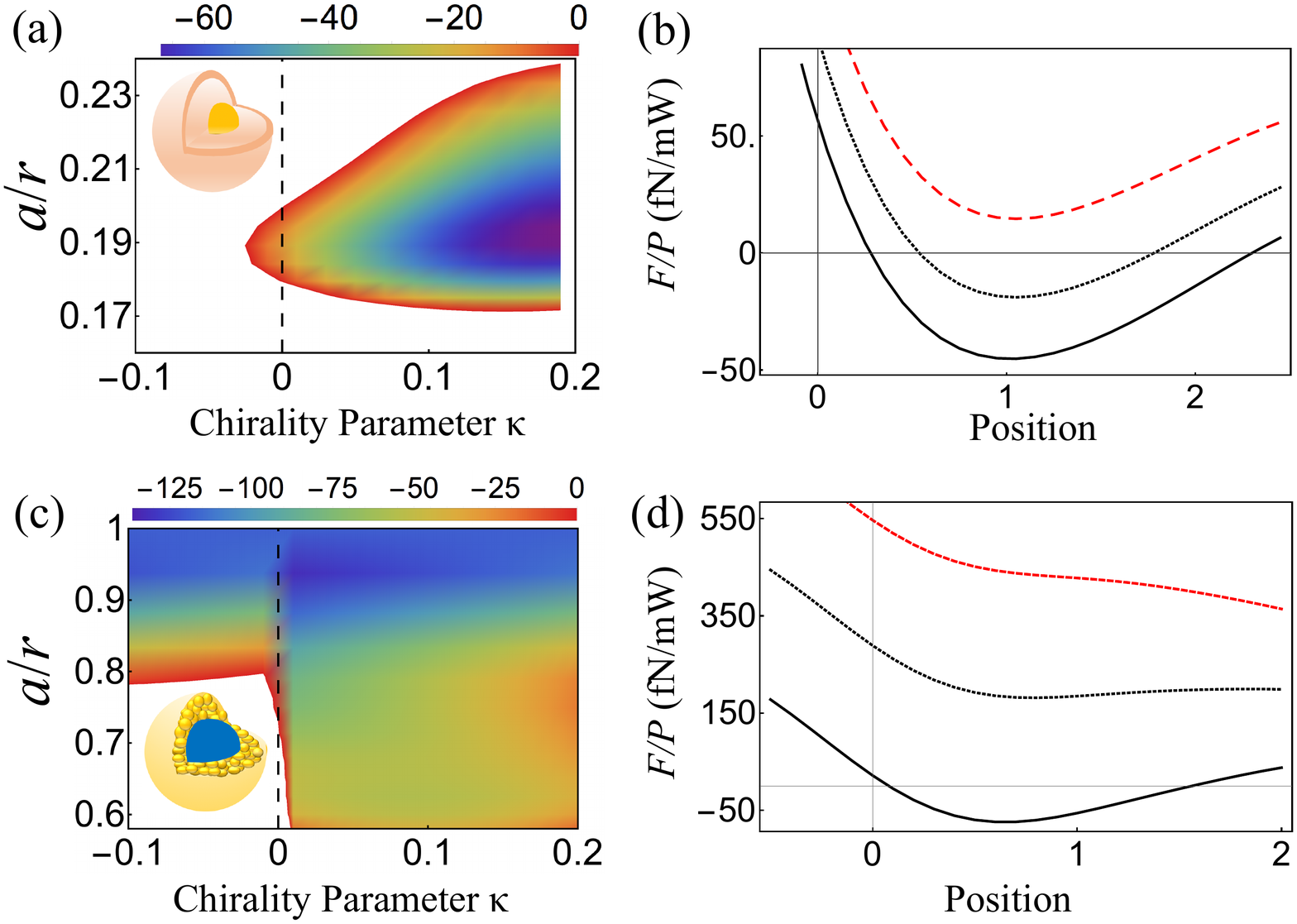}
\caption{(a,c) Density plot  
showing the range of chirality parameter $\kappa$ and aspect ratio $a/r$ (see Fig.~\ref{setup}) leading to negative axial forces 
 at the position (a) $z/r=1.3$  and  (c) $z/r=0.5$. The color scale represents the force in femtonewtons for a laser beam power of  $1\,{\rm mW}.$
 (b,d) Axial force as a function of position (in units of the sphere outer radius $r$)
for (b) $\kappa=-0.05$  (dash),  $\kappa=0$ (dot), $\kappa=0.05$  (solid)
and (d) $\kappa=-0.05-0.007\,{\rm i}$  (dash),  $\kappa=0$ (dot), $\kappa=0.05+0.007\,{\rm i}$ (solid). 
Panels (a,b) refer to  metal-organic core-shell particles  made of a gold core ($a=70\,{\rm  nm}$) coated with a dielectric chiral shell,
whereas (c,d) correspond to  raspberry-like core-satellite particles
 made of several layers of metallic nanospheres randomly distributed around a silica core
 (radius  $a=300\,{\rm nm}$).
The outer radius is $r= 370\,{\rm nm}$ and $450\,{\rm nm}$ in panels (b) and (d), respectively. 
 In all cases we take a left-handed circularly-polarized laser beam. } 
 \label{Axial_Force}
\end{figure}

 For all numerical examples discussed below, we take the typical values \cite{Dutra2014}
  $\lambda_0=1064\,{\rm nm},$ 
 $n_w=1.332,$ $n_g= 1.51,$
   $\gamma=1.226$ and   NA$=1.3$. 
   We take right-handed circular polarization ($\sigma=-1$), but results for the opposite helicity $\sigma=1$ can be trivially obtained from those shown here by changing the signs of both $\alpha$ and $\kappa.$ 
   We consider different realistic examples of core-shell nanospheres of both technological and scientific interest. 
   We start with metal-organic nanoparticles~\cite{wu2015} consisting of a gold core (radius $a=70\,{\rm nm}$) and a dielectric chiral shell with a relative permittivity $\epsilon_s=2.89.$ 
Fig.~\ref{Axial_Force}(a) presents a density plot showing the negative (backward) values of the axial force $F_z,$
required for trapping, 
 as function of the chirality parameter $\kappa$ and the aspect ratio $a/r$
  ($r=$  outer radius). Numerical values are given for the force divided by the laser beam power $P$ in the sample region. We take the sphere center at the plane $z/r=1.3,$  where $z=0$ corresponds to the paraxial focal plane. 
  Outside the colored region shown in Fig.~\ref{Axial_Force}(a), radiation pressure dominates and the optical force is positive~\cite{Ashkin1992}, thus expelling the particle  from the focal region. 
  Thus, we select  right-handed (left-handed) shells  by taking $\sigma=-1$ (+1) trapping beams. 

  To confirm chiral resolution and have further insight into the trapping conditions, we plot in Fig.~\ref{Axial_Force}(b) the variation of $F_z/P$ with the axial position of the sphere in units of the outer radius
  $r=370\,{\rm nm}.$
The particle with $\kappa=0.05$ (solid) is indeed trapped at a stable equilibrium position close to the focus ($z=0$), while the force for the opposite value of $\kappa$ is positive everywhere 
along the beam axis (dashed), thus flushing the particles with the  opposite handedness  away from the focal region. 
 For achiral particles (dotted), the optical axial force lies in between the two previous cases, and a less stable trapping is achieved further away from the focal point. 
 
 In Figs.~\ref{Axial_Force}(c,d) we consider a second example of 
 plasmonic particle, composed of randomly distributed metallic nanoparticles attached to a dielectric spherical core. Such 
  so-called plasmonic ``raspberries" are dielectric spheres decorated with metallic nanoparticles that show high optical magnetism~\cite{Muhlig2011,Sheikholeslami2013,Ponsinet2015,Qian2015,Wu2017}. 
  Since disorder is known to produce geometries lacking mirror symmetry \cite{pinheiro2017}, we discuss here how to characterize the chirality of an individual plasmonic raspberry by 
 employing optical tweezers. 
 We model  these structures
   as core-shell composites with a silica spherical core (radius $a=300\,{\rm nm}$) coated with homogeneous chiral metallic shells (permittivity 
 $\epsilon_s=2.89+0.03\,{\rm i}$). 
In Fig.~\ref{Axial_Force}(c), we  show that 
trapping is enantioselective when the shell thickness $t$ is larger than $\sim 100\,{\rm nm}.$ Those length scales are  similar to the 
size parameters of the multi-layered raspberries reported in Ref.~\cite{Qian2015}. 
For a shell thickness $t=150\,{\rm nm},$ 
Fig.~\ref{Axial_Force}(d) shows that 
particles with $\kappa=0.050+0.007\,{\rm i}$ are trapped (solid), whereas achiral (dotted) or chiral with opposite 
handedness (dashed) are pushed away from the focal region by radiation pressure.

\begin{figure}
\includegraphics[width = 3. in]{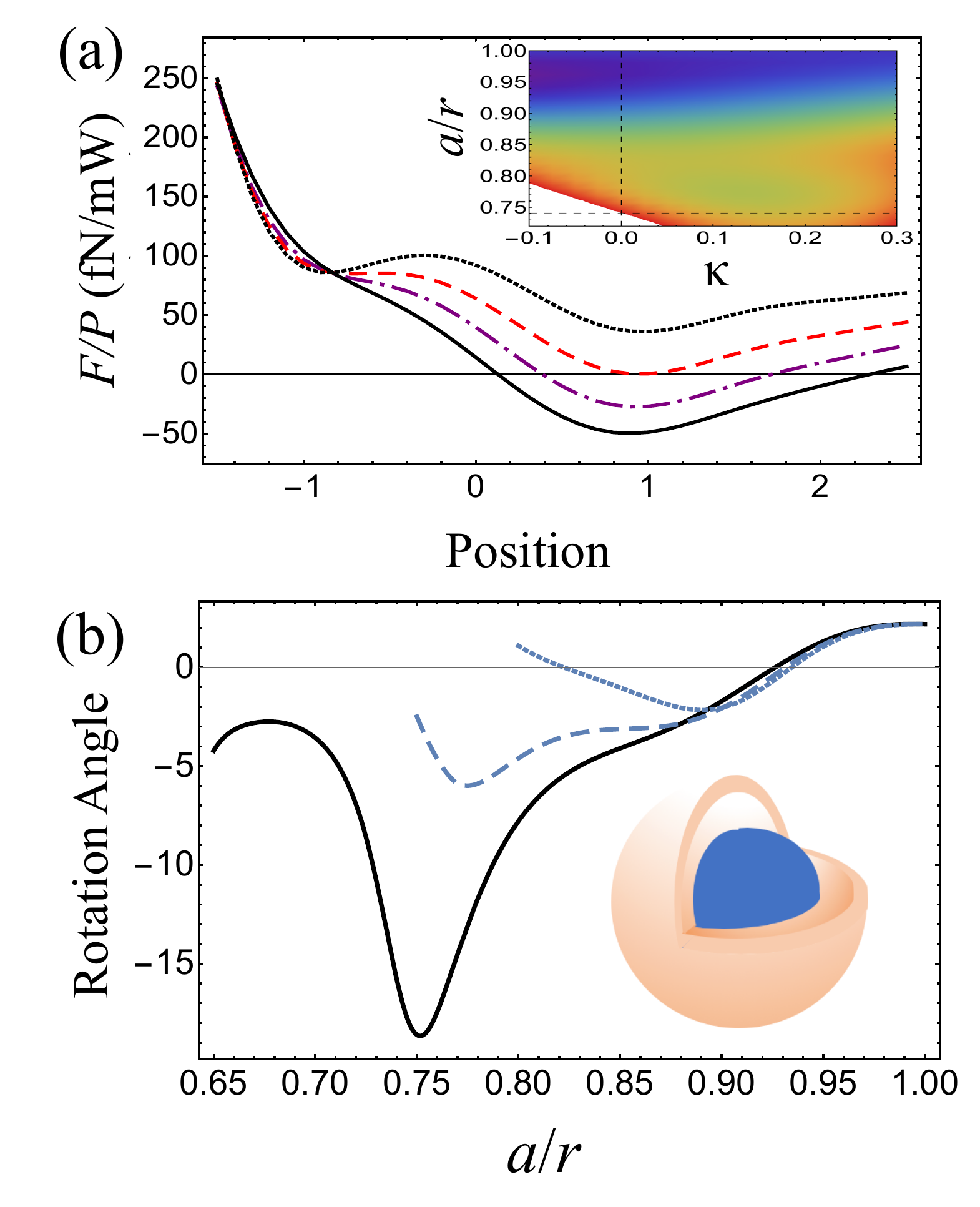} 
\caption{(a)  Axial force as a function of position (in units of the sphere outer radius $r$) for
chirality parameter  $ \kappa=-0.05$  (dot),  $\kappa=0$ (dash), $\kappa=+0.05$ (dot-dash) and $\kappa=0.12$ (solid).
Inset:
density plot  showing the range of $\kappa$ and aspect ratio $a/r$ leading to negative axial forces required for trapping,
 at the fixed nanosphere position $z/r=1.$
(b) Rotation angle $\alpha$ of the particle equilibrium position as a function of aspect ratio $a/r,$
for  chirality parameters  $\kappa= 0.12$ (solid), $\kappa=0$ (dash) and $\kappa= - 0.1$ (dot). 
For the last two cases, the curves end at their respective thresholds of no trapping, since radiation pressure dominates for smaller values of $a/r.$
In all panels, we consider a silica core of radius
 $a=300\,{\rm nm}$ and a dielectric chiral shell with $\epsilon_s=2.89.$ The shell thickness is $t=375\,{\rm nm}$ in the main plot of panel (a). }
\label{Dielectric}
\end{figure}

Our proposal can also be applied outside the domain of chiral plasmonics, as   demonstrated in Fig.~\ref{Dielectric}.
We consider a core-shell dielectric particle with a silica core of radius $a=300\,{\rm nm}$ and a dielectric chiral shell of 
permittivity $\epsilon = 2.89$ and
thickness $t.$
The axial force variation with position is shown in the main plot of panel (a)
  for $t=175\,{\rm nm}.$ Stable trapping is achieved for 
   realistic positive values for the chirality parameter $\kappa$~\cite{dionnereply},
 while nanospheres coated with the opposite handedness  are expelled from the focal region as in our previous examples. 
A more complete picture is provided by the 
 density plot in the
inset of \ref{Dielectric}(a),  where the conventions are the same as in Figs.~\ref{Axial_Force}(a,c). 
Enantioselectivity is  possible in the range $100\,{\rm nm} < t < 200\,{\rm nm},$
 whereas spheres coated with thin layers ($t<100\,{\rm nm}$) are trapped regardless of the sign of $\kappa$, recovering the results for achiral homogeneous nanospheres \cite{MaiaNeto2000,Mazolli2003} as $t\rightarrow 0.$

The range of enantioselection can be further extended into the 
 thickness interval $40\,{\rm nm}\stackrel{<}{\scriptscriptstyle\sim} t \stackrel{<}{\scriptscriptstyle\sim}100\,{\rm nm}$  by driving the sample laterally so as to displace the trapped particle away from the beam axis, as discussed in connection with
Fig.~\ref{setup}. Due to the transfer of optical angular momentum (OAM), the equilibrium position is rotated by an angle $\alpha$ 
around the beam axis. 
In Fig.~\ref{Dielectric}(b), we plot $\alpha$ as a function of the aspect ratio $a/r.$ 
For thin shells corresponding to aspect ratios in the range $0.88\stackrel{>}{\scriptscriptstyle\sim} a/r
\stackrel{>}{\scriptscriptstyle\sim}0.74,$ the curves corresponding to different chiralities are sufficiently far apart to distinguish between shells with positive, negative and vanishing 
values of $\kappa$ shown in Fig.~\ref{Dielectric}.

For homogenous achiral spheres,
 the rotation angle $\alpha$ is usually opposite to the polarization handedness of the trapping beam~\cite{Diniz2019} (so-called negative optical torque~\cite{Chen2014,Hakobyan2014,Hakobyan2015}).
 Indeed, 
 we find a positive $\alpha$  in the limit  $a/r\rightarrow 1$
 when taking right-handed circular polarization (negative spin angular momentum) as 
 in the example of Fig.~\ref{Dielectric}(b). On the other hand, Fig.~\ref{Dielectric} shows 
 that optical torque goes positive when considering thicker chiral shells. 
 
The solid line, corresponding to $\kappa= 0.12,$ shows that 
the rotation angle depends strongly on the shell thickness and has a peak value $\alpha\approx 19^{\rm o}$ close to $t= 100\,{\rm nm}.$
This is $\sim 5$ times larger than the typical values predicted and measured for achiral homogeneous particles of similar dimensions~\cite{Diniz2019}. 
Such an  impressive increase makes trapped single spheres with chiral coatings  
particularly suitable for experimental demonstrations of optical torque in optical tweezers.

\begin{figure}
\includegraphics[width = 3.4 in]{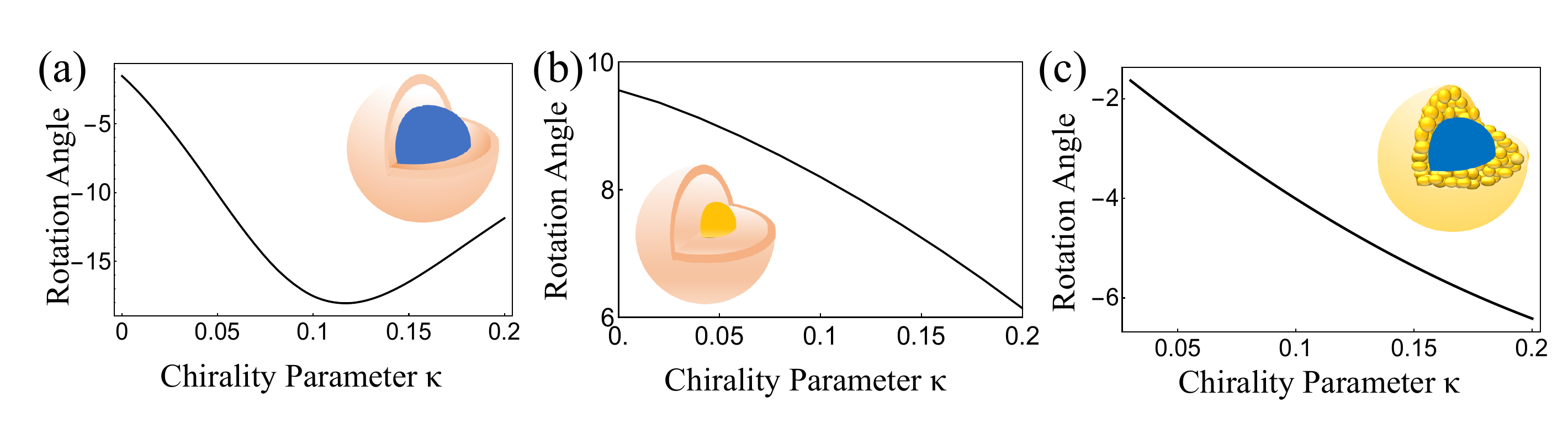} 
\caption{Rotation angle $\alpha$ of the particle equilibrium position as a function of the chirality parameter 
$\kappa$ for (a)  dielectric core-shell,
(b)  metal-organic core-shell and (c) 
 raspberry-like core-satellite particles. The geometrical dimensions in (a,b,c) are the same as in Figs.~\ref{Dielectric}, \ref{Axial_Force}(b) and \ref{Axial_Force}(d),
respectively.
As in the previous figures, we take left-handed circular polarization.
In panel (c), we take ${\rm Im}(\kappa)=0.007.$}
 \label{Azimuthal_Force_2}
\end{figure}

The results of Fig.~\ref{Dielectric}(b)  suggest that the transfer of OAM in
 optical tweezers might provide an efficient method for characterizing the thickness of chiral coatings. 
 In fact, the rotation angle also depends strongly on the chirality parameter $\kappa$ and can be employed to characterize
 its value for individual particles with a high resolution. 
In Fig.~\ref{Azimuthal_Force_2},  we plot the rotation
 angle $\alpha$ as a function of the chirality parameter $\kappa$
 for the three examples we have considered in this letter: (a) a dielectric core-shell with a silica core of radius $a=300\,{\rm nm}$ and a chiral shell of thickness
 $t=375\,{\rm nm};$ (b) a metal-organic particle with a gold core of radius $a=70\,{\rm nm}$ and chiral dielectric shell 
 of thickness
 $t=300\,{\rm nm};$ and (c) a plasmonic raspberry with a silica core of radius $a=300\,{\rm nm}$ and a chiral metallic shell of thickness  $t=150\,{\rm nm}.$
 Remarkably, the optical torque is positive 
 for the dielectric and plasmonic raspberry particles, and negative for the metal-organic one. 
 For all three cases,  $\alpha$ displays a strong, approximately linear dependence in the  
 range $0\le \kappa \stackrel{<}{\scriptscriptstyle\sim} 0.1,$
 allowing for a good resolution in the determination of $\kappa$ in this important range.
 Given a typical precision $\delta\alpha\sim 0.2^{\rm o}$ in the determination of the rotation angle \cite{Diniz2019}, 
 we estimate 
 from Fig.~\ref{Azimuthal_Force_2}
  a chirality resolution $\delta\kappa\sim 10^{-3}$ for the dielectric core-shell particle, and $\delta\kappa\sim 10^{-2}$
 for the 
 metal-organic and raspberry composites. 
 
 As discussed previously, 
 it is also possible to 
 trap nanospheres coated with the opposite handedness by
  switching to left-handed circular polarization.
  In this case, the sign of the rotation angle is 
  reversed, and the resulting variation allows for the characterization of negative values of $\kappa.$ 
  
In conclusion, 
we have shown that optical tweezing of chiral core-shell nanospheres  is  enantioselective, 
since the handedness
 allowing for trapping can be selected by choosing the appropriate circular polarization of the laser beam.
 For the plasmonic raspberry-like and the core-shell dielectric particles, enantioselection is achieved 
as long as 
the chiral shell thickness exceeds $\sim 100\,{\rm nm}.$ Nanospheres with the opposite handedness are simply pushed away from the 
focal region by radiation pressure. 
In addition, we put forward 
a method for characterization of the chirality parameter $\kappa$ 
 based on the rotation of the trapped particle equilibrium position when it is displaced off-axis by 
driving the sample laterally. 

We generally find large rotation angles that
should be more easily measurable than in the experiment reported in Ref.~\cite{Diniz2019}.
 The angle is strongly dependent on the chirality parameter, thus providing 
a good resolution for its characterization in the range  $ |\kappa|\stackrel{<}{\scriptscriptstyle\sim} 0.1.$  
Our method is particularly suited to unveil chirality of plasmonic raspberries, which is expected to exist due to the random arrangement of metallic satellites that lacks mirror symmetry.
In contrast to traditional methods to probe chirality, such as circular dichroism and rotatory power, here  
the chiral optical response does not show up as the average response of particles in suspension, but  
only when probing  individual particles, each one with its unique handedness.
Altogether our results demonstrate a twofold mechanism for all-optical manipulation of chiral nanoparticles, 
based on optical trapping and rotation, 
 paving the way for the investigation 
 of the chiral response of plasmonic nanoparticles and applications. 

\begin{acknowledgements}
We thank K. Diniz, D. S. Ether Jr, L. B. Pires and N. B. Viana for inspiring discussions.  This work has been supported by 
 the Brazilian agencies National Council for Scientific and Technological Development (CNPq), 
  Coordination for the Improvement of Higher Education Personnel (CAPES),  the National Institute of Science and Technology Complex Fluids (INCT-FCx),
and the Research Foundations of the States of Minas Gerais (FAPEMIG), Rio de Janeiro (FAPERJ) and S\~ao Paulo (FAPESP).
F. A. P. thanks the Royal Society -- Newton Advanced Fellowship (grant no. NA150208) for financial support. 
 \newpage
\newpage
\end{acknowledgements}

%\newpage\newpage\newpage
\hbox{}\newpage

\include{supplement}

\onecolumngrid
%\onecolumn 
\begin{center}
\textbf{\large Supplemental Materials: Enantioselective manipulation of chiral nanoparticles using optical tweezers }
\end{center}
\twocolumngrid

 This supplement contains an expanded theoretical description of optical tweezers of particles with a chiral shell.

We consider a circularly-polarized Gaussian beam at the objective entrance port, with $\sigma = \pm 1$ denoting lefthanded/righthanded polarization. 
The dimensionless optical force efficiency (see main text for definition in terms of the optical force)  is written as 
\begin{equation}
{\bf Q}_{}(\rho,\phi,z) = {\bf Q}_{s}(\rho,\phi,z)  +{\bf Q}_{e}(\rho,\phi,z).
\end{equation} 
The extinction term  ${\bf Q}_e$  accounts for  the rate of momentum removal from the incident beam, while the scattering  contribution  ${\bf Q}_s $ 
corresponds to the negative of the rate of momentum carried away by the scattered field.  
The explicit expressions for their cylindrical components are given below as  sums over multipoles of the  form
\[
\sum_{\ell m}(...) \equiv \sum_{\ell=1}^{\infty}\sum_{m=-\ell}^{\ell}(...)
\]

\begin{widetext}

\begin{itemize}

\item Scattering axial component
\[
Q_{s z}(\rho,\phi,z) = -\frac{8\gamma^2}{AN}{\rm Re}\sum_{\ell m}\frac{\sqrt{\ell(\ell+2)(\ell+m+1)(\ell-m+1)}}{\ell+1} \nonumber\\
  \biggl[(A_{\ell}A_{\ell+1}^{*}+B_{\ell}B_{\ell+1}^{*})
G^{(\sigma)}_{\ell,m}G^{(\sigma)*}_{\ell+1,m}\biggr] 
\]\begin{eqnarray*}
 -\frac{8\gamma^2}{AN}\sigma{\rm Re}\sum_{\ell m}
\frac{(2\ell+1)}{\ell(\ell+1)} m A_{\ell}B_{\ell}^{*}\vert G^{(\sigma)}_{\ell,m}\vert^2
\end{eqnarray*} 

\item Scattering radial component

\[
Q_{s\rho}(\rho,\phi,z) = \frac{4\gamma^2}{AN}\sum_{\ell m}\frac{\sqrt{\ell(\ell+2)(\ell+m+1)(\ell+m+2)}}{\ell+1}\\
 {\rm Im}\biggl\lbrace (A_{\ell}A_{\ell+1}^{*}+B_{\ell}B_{\ell+1}^{*})\]\[
\left[ G^{(\sigma)}_{\ell,m}G^{(\sigma)*}_{\ell+1,m+1}+ G^{(\sigma)}_{\ell,-m}G^{(\sigma)*}_{\ell+1,-(m+1)}\right]\biggr\rbrace
 -\frac{8\gamma^2}{AN}\sigma\sum_{\ell m} \frac{(2\ell+1)}{\ell(\ell+1)}\sqrt{(\ell-m)(\ell+m+1)}   \biggl[{\rm Re}(A_{\ell}B_{\ell}^{*})
{\rm Im}( G^{(\sigma)}_{\ell,m}G^{(\sigma)*}_{\ell,m+1}) \biggl] \]
\item Scattering azimuthal component

\[
Q_{s\phi}(\rho,\phi,z) = -\frac{4\gamma^2}{AN}\sum_{\ell m}\frac{\sqrt{\ell(\ell+2)(\ell+m+1)(\ell+m+2)}}{\ell+1}\\
 {\rm Re}\biggl\lbrace (A_{\ell}A_{\ell+1}^{*}+B_{\ell}B_{\ell+1}^{*})\times\]\[
\left[ G^{(\sigma)}_{\ell,m}G^{(\sigma)*}_{\ell+1,m+1}-G^{(\sigma)}_{\ell,-m}G^{(\sigma)*}_{\ell+1,-(m+1)}\right]\biggr\rbrace
+\frac{8\gamma^2}{AN}\sigma\sum_{\ell m}
 \frac{(2\ell+1)}{\ell(\ell+1)}\sqrt{(\ell-m)(\ell+m+1)} \times {\rm Re}(A_{\ell}B_{\ell}^{*})
{\rm Re}( G^{(\sigma)}_{\ell,m}G^{(\sigma)*}_{\ell,m+1})\]

\item Extinction axial component

\begin{equation}
Q_{ez}(\rho,\phi,z)=\frac{4\gamma^2}{AN}{\rm Re}\sum_{\ell m}(2\ell+1)
G^{(\sigma)}_{\ell,m}
\left[(A_{\ell}+B_{\ell})
G'^{(\sigma)*}_{\ell,m}
\right],
\end{equation}

\item Extinction radial component

\[
Q_{e\rho}(\rho,\phi,z)=\frac{2\gamma^2} {AN}{\rm Im}\sum_{\ell m}(2\ell+1)G^{(\sigma)}_{\ell,m}
\biggl[(A_{\ell}+B_{\ell}) 
\left(G^{(\sigma)-}_{\ell ,m+1} - G^{(\sigma)+}_{\ell,m-1}\right)^*\biggr]\]
\item Extinction azimuthal component

\[
Q_{e\phi}(\rho,\phi,z)=-\frac{2\gamma^2} {AN}{\rm Re}\sum_{\ell m}(2\ell+1)G^{(\sigma)}_{\ell,m}
\biggl[(A_{\ell}+B_{\ell}) 
\left(G^{(\sigma)+}_{\ell,m-1}+G^{(\sigma)-}_{\ell,m+1}\right)^*\biggr]\]
\end{itemize}

In addition to the multipole coefficients $G_{\ell m}^{(\sigma)}(\rho,\phi,z)$ defined in Eq.~(3) of the main letter, we also define 
\begin{equation*}
G'^{(\sigma)}_{\ell,m}(\rho,\phi,z)=\int_{0}^{\theta_0}d\theta\sin\theta \cos\theta_{w}  \sqrt{\cos \theta}\,T(\theta)\,e^{-\gamma^2\sin^2\theta}d_{m,\sigma}^{\ell}(\theta_{w})\, J_{m-\sigma}\left( k\rho \sin\theta\right)  e^{i[\Phi_{w}(\theta)+k_{w}\cos\theta_{w} z ]}, \label{multipole coefficient}
\end{equation*}
\begin{equation*}
G_{\ell,m}^{(\sigma)\pm}(\rho,\phi,z)=\int_{0}^{\theta_0}d\theta_{}\sin\theta \sin\theta_{w} \sqrt{\cos \theta}\,T(\theta)\,e^{-\gamma^2\sin^2\theta}d_{m\pm 1,\sigma}^{\ell}(\theta_{w})\, J_{m-\sigma}\left( k \rho\sin\theta \right)     e^{i[\Phi_{w}(\theta)+k_{w}\cos\theta_{w} z ]},
\end{equation*}
\end{widetext}

with  $\theta_{w}=\sin^{-1}(\sin\theta/N)$  and $N= n_{w}/n_g.$
The phase $\Phi_{w}$ accounts for the spherical aberration arising from the refractive index mismatch at the glass-water interface:
\begin{equation*}
\Phi_{w}(\theta)=k\left( -L/N\cos\theta+NL\cos\theta_{w}\right), \label{aberration_interface}
\end{equation*}
where  $L$ represents  the distance between the paraxial focal plane and the  glass slide.
We also take 
   $T(\theta)= \frac{2\cos\theta}{\cos\theta+N\cos\theta_{w}}$ for the Fresnel transmission amplitude (neglecting polarization dependence since $N\approx 1$). 

 The factor 
\[
A = 16\gamma^2\int_0^{s_0} ds\,s \exp(-2\gamma^2s^2)\frac{\sqrt{(1-s^2)(N^2-s^2)}}{(\sqrt{1-s^2}+\sqrt{N^2-s^2})^2}
\]
 is  the fraction of beam power transmitted into the sample chamber, with $s_0=\mbox{NA}/n_g.$

The  optical force components also depend on the effective external Mie coefficients  $A{}_{\ell}$ and $B{}_{\ell}$  
which we derive as follows:
 \begin{equation}
\begin{aligned}
 A_{\ell}= a_{\ell}+ i\sigma d_{\ell} \\B_{\ell}= b_{\ell} - i\sigma c_{\ell}.
 \end{aligned} \label{Mie-cof}
\end{equation}   
$A_{\ell}$ and $B_{\ell}$ 
are convenient for a number of applications involving circularly-polarized fields scattered by chiral media. 
The coefficients $ a_{\ell} $,  $ b_{\ell} $,  $ c_{\ell}$  and  $ d_{\ell} $ are the external Mie  coefficients for a sphere made of chiral material \cite{BohrenHuffman}. 
 For a core-shell nanosphere, their explicit expressions \cite{Bohren1975} are 
 given in terms of the size parameters $\alpha=k_w a$ and $v=k_w r$ corresponding to the core and outer radii $a$ and $r$.  
 The refractive indexes of the chiral shell (with respect to the host) are $N_{L/R}= (\sqrt{\epsilon\mu}\pm \kappa)/n_w,$ where $\kappa$ is the chirality parameter (see main text).  We also need the relative refractive index of the core $N_{\rm I}$ with respect to the host medium of index $n_w.$ Finally, $N_{\rm II}= (N_L+N_R)/2$ is the average 
 relative index of the chiral shell. 

 \begin{eqnarray}
a_{\ell} & = & -\Delta^{-1}_\ell (A_{R\ell} W_{L\ell}+A_{L\ell} W_{R\ell}) \nonumber\\
b_{\ell}& = & -\Delta^{-1}_\ell (B_{L\ell}V_{R\ell}+B_{R\ell} V_{L\ell}) \nonumber\\
c_{\ell} & = & i\Delta^{-1}_\ell (A_{L\ell} V_{R\ell}-A_{R\ell} V_{L\ell} ) \nonumber\\
d_{\ell} & = & i\Delta^{-1}_\ell (B_{R\ell} W_{L\ell}-B_{L\ell} W_{R\ell}) \nonumber
\end{eqnarray}
with 
 \begin{eqnarray}
 \Delta_{\ell} &=&W_{L\ell} V_{R\ell}+B_{L\ell} W_{R\ell} \nonumber\\
A_{ R\ell} &=&X_{ R\ell}(-) \eta_\ell^{(1)} (v) -N_{\scriptstyle \rm II} U_{ R\ell} (-) j_\ell (v) \nonumber \\
A_{L\ell} &=& X_{ L\ell}(+) \eta_\ell^{(1)} (v) -N_{\scriptstyle \rm II} U_{ L\ell} (+) j_\ell(v) \nonumber \\
B_{ L\ell} & = & X_{ L\ell}(-)N_{\scriptstyle \rm II} \eta_\ell^{(1)} (v) - U_{ L\ell} (-) j_\ell (v) \nonumber\\
B_{R\ell} & = & X_{ R\ell}(+) N_{\scriptstyle \rm II}\eta_\ell^{(1)} (v) - U_{ R\ell} (+) j_\ell (v) \nonumber\\
V_{ L\ell} & = & X_{ L\ell}(+) \eta_\ell^{(3)} (v) -N_{\scriptstyle \rm II} U_{ L\ell} (+) h^{(1)}_\ell (v) \nonumber\\
V_{R\ell} & = & X_{ R\ell}(-) \eta_\ell^{(3)} (v) -N_{\scriptstyle \rm II} U_{ R\ell} (-) h^{(1)}_\ell (v) \nonumber\\
W_{ L\ell} &=&X_{ L\ell}(-)N_{\scriptstyle \rm II} \eta_\ell^{(3)} (v) - U_{ L\ell} (-) h^{(1)}_\ell (v) \nonumber\\
W_{R\ell} & = & X_{ R\ell}(+) N_{\scriptstyle \rm II}\eta_\ell^{(3)} (v) - U_{ R\ell} (+) h^{(1)}_\ell (v) \nonumber
\end{eqnarray}
We have also introduced the functions 
\begin{eqnarray}
X_{R\ell}(\pm)&=& j_\ell (N_R v)+D_{4\ell} y_\ell(N_R v)\pm D_{2\ell} y_\ell(N_L v) \nonumber\\
X_{L\ell}(\pm)&= &j_\ell(N_L v)+D_{1\ell}y_\ell(N_L v)\pm D_{3\ell} y_\ell(N_R v) \nonumber\\
U_{R\ell}(\pm)&= &\eta^{(1)}_\ell(N_R v)+D_{4\ell}\eta^{(2)}_\ell(N_R v)\pm D_{2\ell}\eta^{(2)}_\ell(N_L v) \nonumber\\
U_{L\ell}(\pm)&= &\eta^{(1)}_\ell(N_L v)+D_{1\ell}\eta^{(2)}_\ell(N_L v)\pm D_{3\ell}\eta^{(2)}_\ell(N_R v) \nonumber
\end{eqnarray}
where 
\begin{eqnarray}
D_{1\ell}&=& -\Delta^{-1}_{\ell}[G_\ell(N_R)H_\ell(N_L)+F_\ell(N_R)K_\ell(N_L)]\nonumber\\
D_{2\ell}&= &\Delta^{-1}_{\ell}[F_\ell(N_R)K_\ell(N_R)-G_\ell(N_R)H_\ell(N_R)]\nonumber\\
D_{3\ell}&= &\Delta^{-1}_{\ell}[G_\ell(N_L)H_\ell(N_L)-F_\ell(N_L)K_\ell(N_L)]\nonumber\\
D_{4\ell}&=&- \Delta^{-1}_{\ell}[G_\ell(N_L)H_\ell(N_R)+F_\ell(N_L)K_\ell(N_R)]\nonumber
\nonumber
\end{eqnarray}

$F,G,H$ and $K$ are functions of the refractive index variable $N=N_L,N_R$ defined as
\begin{eqnarray}
F_{\ell}(N)&=&N_{\scriptstyle \rm II}y_\ell {(N\alpha)}\eta^{(1)}_\ell {(N_{\rm I}\alpha)}-N_{\rm I}\eta^{(2)}_\ell {(N \alpha)}j_\ell {(N_{\rm I}\alpha)}\nonumber\\
G_{\ell} (N) &=&N_{\rm I}y_\ell {(N\alpha)}\eta^{(1)}_\ell {(N_{\rm I}\alpha)}-N_{\scriptstyle \rm II}\eta^{(2)}_\ell {(N \alpha)}j_\ell {(N_{\rm I}\alpha)}\nonumber\\
H_{\ell} (N) &=&N_{\rm I}j_\ell {(N\alpha)}\eta^{(1)}_\ell {(N_{\rm I}\alpha)}-N_{\rm I}\eta^{(1)}_\ell {(N \alpha)}j_\ell {(N_{\rm I}\alpha)}\nonumber\\
K_{\ell}(N) &=&N_{\scriptstyle \rm II}j_\ell {(N\alpha)}\eta^{(1)}_\ell {(N_{\rm I}\alpha)}-N_{\scriptstyle \rm II}\eta^{(1)}_\ell {(N \alpha)}j_\ell {(N_{\rm I}\alpha)}\nonumber
\end{eqnarray}

$ j_{\ell}(\rho)  $, $ y_{\ell}(\rho) $ are the spherical Bessel functions of the first and second kind, respectively; whereas
  $ h^{(1)}_{\ell} (\rho)$ is the spherical Hankel function of  the first  kind \cite{DLMF25.12}. We also define
\begin{eqnarray}
\eta^{(1)}_\ell (\rho)&=& \frac{1}{\rho}d[\rho j_\ell(\rho)]/d\rho \nonumber\\ 
\eta^{(2)}_\ell (\rho)&= &\frac{1}{\rho}d[\rho y_\ell(\rho)]/d\rho \nonumber \\
\eta^{(3)}_\ell (\rho)&= &\frac{1}{\rho}d[\rho h_\ell^{(1)}{}(\rho)]/d\rho  \nonumber
\end{eqnarray}

\end{document}